\documentclass[12pt,twocolumns]{iopart}
\usepackage{iopams}
\usepackage{fancyhdr}
\usepackage{float}
\usepackage{graphicx}
\usepackage{graphics}
\usepackage{amsthm}
\usepackage{color}
\usepackage{hyperref}
\usepackage[update,prepend]{epstopdf}
\DeclareGraphicsExtensions{.pdf,.jpg,.png,.eps, .ps}
\usepackage{textcomp}

\newcommand{\HFCO}{Ho$_2$FeCoO$_6$}
\newcommand{\EFCO}{Eu$_2$FeCoO$_6$}
\newcommand{\HFO}{HoFeO$_{3}$}
\newcommand{\HCO}{HoCoO$_{3}$}

\newcommand{\THo}{$T_\mathrm{Ho}$}

\begin{document}
%
%
%
\title[Spin reorientation and disordered rare earth magnetism in Ho$_2$FeCoO$_6$]{Spin reorientation and disordered rare earth magnetism in Ho$_2$FeCoO$_6$}
\author{Haripriya G. R.$^{1}$, Harikrishnan S. Nair$^2$, Pradheesh R.$^{1*}$, S. Rayaprol$^3$, V. Siruguri$^3$, Durgesh Singh$^4$, R. Venkatesh$^4$, V. Ganesan$^4$,  K. Sethupathi $^{1}$, V. Sankaranarayanan$^1$ }
\address{$^1$Low Temperature Physics Laboratory, Department of Physics, Indian Institute of Technology Madras, Chennai-600036, India}
\address{$^2$ Department of Physics, 500 West University Ave, University of Texas at El Paso, TX 79968, USA}
\address{$^3$ UGC-DAE Consortium for Scientific Research - Mumbai Centre, R-5 Shed, BARC Campus, Mumbai-400085, India}
\address{$^4$Low Temperature Laboratory, UGC-DAE Consortium for Scientific Research, University Campus, Khandwa Road, Indore-452001, India}
\address{$*$Current address: Department of Physics, National Institute of Technology Calicut, Kozhikode - 673601, Kerala, India}
\ead{ksethu@iitm.ac.in, haripriya@physics.iitm.ac.in}
%
%
\clearpage
\begin{abstract}
We report the experimental observation of spin reorientation in the double perovskite \HFCO. 
The magnetic phase transitions in this compound are characterized and studied through magnetization and specific heat, and the magnetic structures are elucidated through neutron powder diffraction.
Two magnetic phase transitions are observed in this compound $\textendash$ one at $T_\mathrm{N1} \approx$ 250~K, from paramagnetic to antiferromagnetic, and the other at $T_\mathrm{N2} \approx$ 45~K, from a phase with mixed magnetic structures to a single phase through a spin reorientation process. The magnetic structure in the temperature range 200~K $\textendash$ 45~K is a mixed phase of the irreducible representations $\Gamma_1$ and $\Gamma_3$, both of which are antiferromagnetic. The phase with mixed magnetic structures that exists in \HFCO\ gives rise to a large thermal hysteresis in magnetization that extends from 200~K down to the spin reorientation temperature. At $T_\mathrm{N2}$, the magnetic structure transforms to $\Gamma_1$.
Though long-range magnetic order is established in the transition metal lattice, it is seen that only short-range magnetic order prevails in Ho$^{3+}\textendash$ lattice. Our results should motivate further detailed studies on single crystals in order to explore spin reorientation process, spin switching and the possibility of anisotropic magnetic interactions giving rise to electric polarization in \HFCO.
\end{abstract}
%

%
\submitto{\JPCM}
%
\maketitle
%
%

\section{\label{intro}Introduction}
Rare earth double perovskites of the chemical formula
$R_2BB'O_6$ ($R$ = rare earth, $B/B'$ = transition metal)
are actively investigated in recent times due to their
potential to exhibit multiferroic and multifunctional properties.
The research on double perovskites extends over bulk as well as thin films in order to
realize usable multiferroic device elements
\cite{singh2010multiferroic,yanez2011multiferroic,ramesh2007multiferroics,kobayashi1998room}.
In addition to multiferroicity \cite{yanez2011multiferroic},
observation of metamagnetic steps in magnetization \cite{nair2014magnetization},
exchange bias \cite{nair2015antisite}, 
magnetoresistance and magnetocapacitance \cite{rogado2005magnetocapacitance} 
make these compounds interesting.
Several Mn and Ni $\textendash$ containing double perovskites \cite{booth2009investigation} 
have been studied so far and their magnetic properties are understood
in terms of the degree of cationic ordering and Goodenough-Kanamori (GK)
rules \cite{goodenough1955theory} that predict
ferromagnetism in these structures, or the deviations from GK rules
due to inherent cationic disorder and subsequent departure from ferromagnetism.
Usually, the double perovskite compounds prepared in the laboratory
suffer from the inevitable antisite disorder due to cations interchanging
their crystallographic positions.
The formation of ordered structures demands suitable combination of $R$ and/or $B/B'$ cations
with a proper control over synthesis parameters \cite{sarma2000magnetoresistance}.
Fe-based double perovskites are less-investigated
so far in spite of being an interesting class due to the possibility of exhibiting
high temperature phase transitions. The double perovskites Ho$_2$Fe$B'$O$_6$, 
with a first row transition element at the $B'$-sites have distinct 
properties due to the 3$d$-3$d$ and the 3$d$-4$f$ interactions. 
In the case of $B'$ = Cr, a crystallographically disordered antiferromagnet 
with tuned spin reorientation is found to occur \cite{kotnana2015band} while for 
$B'$ = Mn, a half metallic-ferrimagnet is predicted theoretically \cite{zhang2010first}. 
The Fe-Co combination has its own peculiarities due to their
proximity in the periodic table with similar size and valence states. 
Realizing the potential of multifunctional properties and tunable 
physical properties in $R_2BB'$O$_6$, we carry out an experimental 
investigation of \HFCO\ in the present paper.
\\
The well-studied end members of \HFCO\ are
orthorhombic \HFO\ and \HCO \cite{shao2011single,hcoorthostrain}
which are weak ferromagnets with the rare earth Ho$^{3+}$
ordering magnetically at $\approx$ 4.1~K and 
$\approx$ 3~K respectively \cite{munoz2012magnetic,shao2011single}.
The orthoferrites $R$FeO$_3$ generally have high N\'{e}el
temperatures of the order of $\sim$ 700 K
\cite{marezio1970crystal,eibschutz1967mossbauer}
and in bulk form are not multiferroic though,
recent theoretical calculations on thin films show the emergence
of electric polarization with the application of
strain \cite{zhao2013effect}.
In the orthoferrite \HFO\, spin reorientation of the Fe
moments is reported in the temperature ranges 
50~K $\textendash$ 58~K \cite{shao2011single} and
34~K $\textendash$ 54~K \cite{HFO}, 
caused by Ho$^{3+}$ $\textendash$ Fe$^{3+}$ interactions
and competing Zeeman and van Vleck contribution to the 
anisotropy that initiates the Fe spin reorientation.
When the $B/B'$-site is substituted with other transition
metal cations like Cr or Mn, changes in the spin 
reorientation temperature are observed
\cite{HFCryuan2014tailoring,kotnana2016enhanced}.
In contrast to the orthoferrites, the orthocobaltites 
show lower values of N\'{e}el temperature which correspond to 
Ho ordering. Thus the combination of Co-Fe at the $B/B'$-site can tune the 
N\'{e}el temperatures to far below 700~K but well above 4~K. 
Unlike other orthocobaltites, where the Co\textsuperscript{3+} spins shows 
a transition from high spin (HS) to low spin (LS) via intermediate spin (IS) 
states with decrease in temperature, \HCO\ has its Co\textsuperscript{3+} in LS 
state \cite{munoz2012magnetic}, which is a diamagnetic state.

\section{\label{exptl}Experimental Methods}
Phase-pure polycrystalline powders of \HFCO\ were prepared through citrate
route sol-gel method \cite{haripriya12017temperature}.
The phase purity of the synthesized powders were checked by doing
Rietveld refinement \cite{rietveld_profile_1969} of the powder
x ray diffractograms, recorded using a commercial x ray diffractometer (PANalytical X'Pert Pro),using the FullProf suite of programs \cite{fullprof}.
About 4~g of \HFCO\ was used for neutron powder diffraction experiments
at BARC, Mumbai, India at the powder diffractometer (FCD-PD3), beam line TT-1015 
of the Dhruva reactor using neutrons of wavelength, 1.48~{\AA} \cite{barc_diffractometer}. 
Diffraction patterns were collected at 11 temperature points in the range
2.8~K to 300~K. The data was analysed using FullProf and the
representation analysis of magnetic structures was carried out
using  BasIreps of FullProf suite \cite{rodriguez2011basireps},\cite{ritter2011neutrons} and SARA$h$ \cite{sarah}.
DC magnetic measurements were carried out using a commercial SQUID based VSM (MPMS 3, Quantum Design USA)
in the temperature range 5~K to 375~K, in zero field-cooled warming (ZFC), field-cooled cooling (FCC)
and field-cooled warming (FCW) protocols for various applied fields.
Isothermal magnetization measurements for different temperatures were carried out up to 70~kOe.
AC susceptibility measurements were performed with an ac amplitude of 2.5 Oe, at frequencies 0.3 Hz, 33 Hz, 333 Hz, 666 Hz and 999 Hz.
Specific heat of \HFCO\ was measured in zero field and in applied magnetic fields
10~kOe, 50~kOe and 80~kOe using a commercial Quantum Design PPMS instrument.
\section{\label{results} Results and discussion}
\subsection{\label{mag} Magnetization}
\begin{figure}[!t]
	\centering
	\includegraphics[scale=0.25]{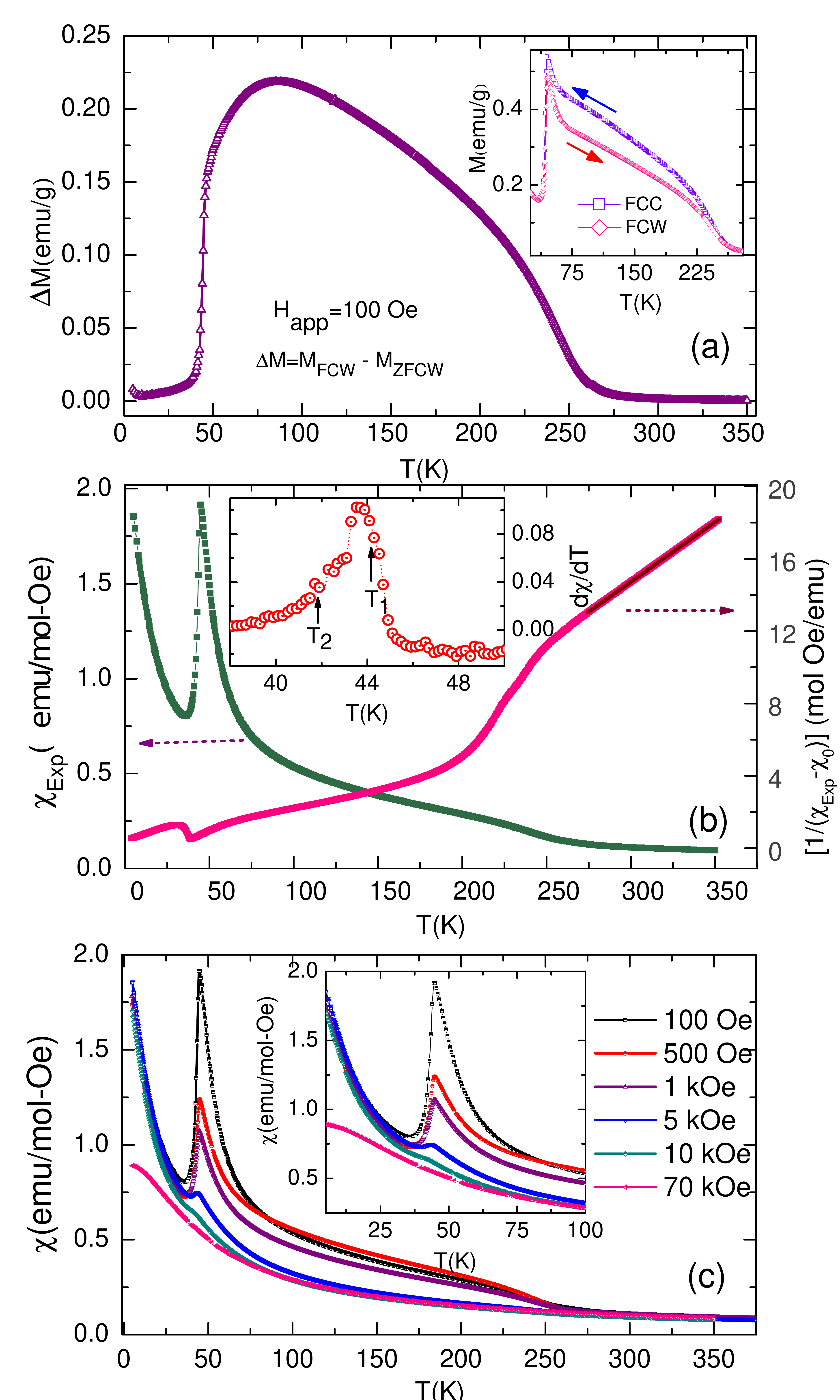}
	\caption{(colour online) (a) The temperature-dependent magnetization of \HFCO\ at an 
		applied field of 100~Oe plotted as $M_\mathrm{FCW} - M_\mathrm{ZFCW}$.
		Magnetic phase transitions occur at $T_\mathrm{N1} \approx$ 250~K and $T_\mathrm{N2} \approx$ 45~K.
		The FCC and FCW arms of magnetization are presented in the inset.
		A large thermal hysteresis over a wide temperature range of 250~K - 50~K.
		(b) The susceptibility $\chi(T)$ = [$\chi_\mathrm{exp} - \chi_\mathrm{0}$] and 
		the corresponding inverse susceptibility at 100 Oe magnetic field.
		The black solid line is the fit according to ideal Curie-Weiss law.
		The inset shows $d\chi/dT$ where a twin-peak is identified at $T_\mathrm{N2}$.
		(c) The effect of applied field on the magnetic phase transitions are indicated. \label{fig:mt}}
\end{figure}

\begin{figure*}[!t]
		\centering
		\includegraphics[scale=0.5]{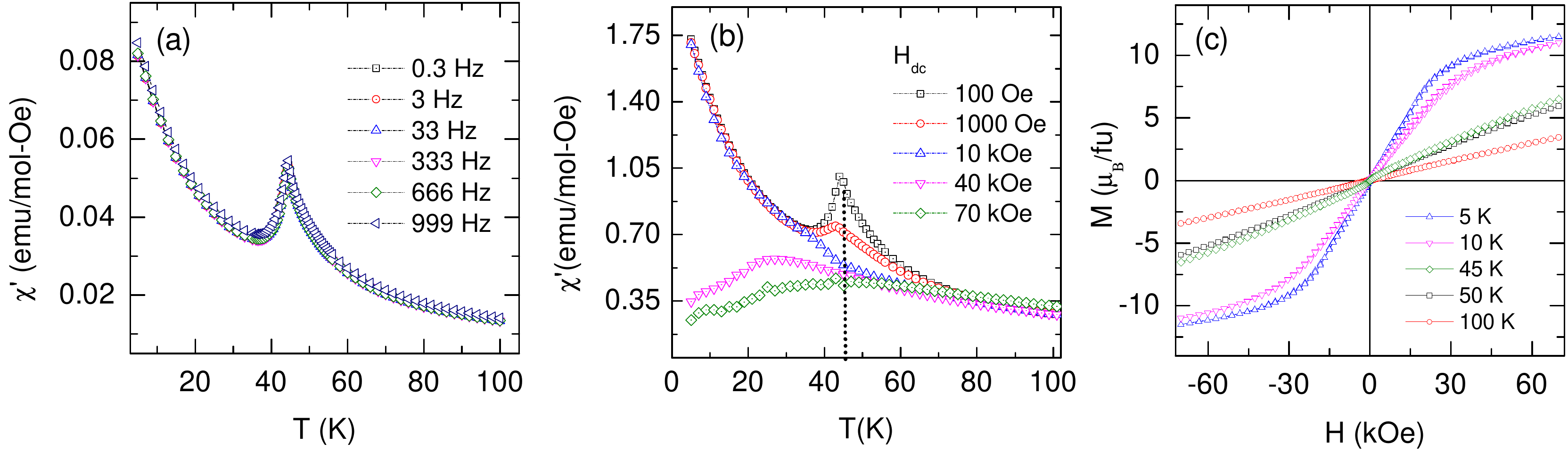}
		\caption{(colour online) (a) No frequency dependence of the real part of ac susceptibility, $\chi'(T)$ over 0.3~Hz to 999~Hz is observed at $T_\mathrm{N2}$ for \HFCO.
			(b) With a superposed DC magnetic field, the peak at $T_\mathrm{N2}$ is observed to get suppressed while slightly shifting to the low temperatures.
			The vertical dash-dotted line is drawn to indicate this shift.
			(c) The isothermal magnetization at different temperatures in the range 5~K to 100~K. \label{fig:acchi}}
\end{figure*}

The temperature dependence of DC magnetization of \HFCO\ at an applied field of
100~Oe was performed following the protocol as ZFC $\rightarrow$ FCC $\rightarrow$ FCW.
In the main panel of figure~\ref{fig:mt} (a) the
field-cooled curves of the magnetization are presented as
$\Delta M$ = [$M_\mathrm{FCW} - M_\mathrm{ZFCW}$], plotted as a function
of temperature.
The FCC and the FCW magnetization curves are presented in the inset of the figure.
A phase transition from paramagnetic to antiferromagnetic phase occurs at
$T_\mathrm{N1} \approx$ 250~K.
Also, a strong thermal hysteresis between the
FCC and the FCW curves begins at $T_\mathrm{N1}$ as can be seen in the inset
of figure~1(a).
A second antiferromagnetic phase transition occurs at $T_\mathrm{N2} \approx$ 45~K where
a peak in magnetization and a subsequent sharp decrease occurs.
Below $T_\mathrm{N2}$, the magnetization tends to increase towards low temperature.
This is attributed to the induced magnetism of Ho$^{3+}$ polarized due to
the ordering in the Fe/Co sub-lattice. A broad peak is seen at
low temperature near 5~K suggesting short-range magnetic
ordering of rare earth moments.
The large region of thermal hysteresis (extended upto about 200~K) is probably
due to the coexistence of symmetry-related
structural phases or due to a multi-domain magnetic structure.
It is interesting to note that orthoferrites of general formula $R$FeO$_3$
($R$ = rare earth), or their derivatives, are known to display spin reorientations and subsequently
multi-domain magnetic structure that can show such hysteresis effects \cite{white1969review,nair_TFMO}.
In fact, a spin-reorientation transition occurs
in the closely related compound \HFO\ signified by twin peaks
at 53~K and 58~K in specific heat \cite{bhattacharjee2002heat,saito2001high}.
It is plausible that the phase transition seen in
\HFCO\ at $T_\mathrm{N2}\approx$ 45~K is similar in nature.
\\
From the presence of Ho in the compound,
we assume time-independent susceptibility contributions
due to diamagnetic and van Vleck terms to the
total magnetic susceptibility of \HFCO.
Hence, a modified Curie-Weiss expression was used
to analyse the magnetic susceptibility:
\begin{equation}
\chi_\mathrm{exp} = \chi_{{o}} + \frac{C}{T - \theta_p}
\label{eqn:1}
\end{equation}
Here, $\chi_{o}$ represents the temperature-independent contributions,
$\theta_p$ is the Curie-Weiss temperature and $C$ is the Curie constant.
In figure~\ref{fig:mt} (b), the magnetic susceptibility
of \HFCO\ is given as $\chi(T)$ = ($\chi_\mathrm{exp} - \chi_o$) on the
left axis and the corresponding inverse susceptibility on the
right axis.
$\chi_o \sim$ 0.0286 emu/mol-Oe was obtained from the fit on
magnetic susceptibility using Eqn~(\ref{eqn:1}).
A linear fit to $1 /(\chi_\mathrm{exp} - \chi_o)$
vs $T$ in the range of 275~K $\textendash$ 350~K
was used to extract the effective paramagnetic moment and the Curie-Weiss temperature.
Thus, an effective magnetic moment, $\mu_\mathrm{eff}$ = $11.74~\mu_\mathrm{B}/$fu 
with a Curie-Weiss temperature
$\theta_{p} = 90(3)$~K were obtained.
The positive value of $\theta_{p}$ suggest predominant ferromagnetic 
interactions.
Interestingly, it can also be noted from the figure~\ref{fig:mt} (b) that extrapolating the
linear fit to low temperatures would indicate the
$1 /(\chi_\mathrm{exp} - \chi_o)$ curve to behave similar to the
Griffiths-like phase seen in double perovskites \cite{nair2011griffiths,liu2014griffiths,chakraborty2016disordered}.
A mean-field approach to estimate the effective magnetic moment and
Curie temperature need not essentially
bring out the details of the complex magnetic interactions
in \HFCO\, as later on we see that prominent spin reorientation
transitions exist in this material.
\\
In the inset of figure~\ref{fig:mt} (b), we plot the derivative $d\chi(T)/dT$
versus temperature to highlight the fact that the transition at
$T_{N2} \approx$ 45~K is in fact a double transition.
From $d\chi(T)/dT$ we identify the two nearby transitions at
$T_1 \approx$ 44.5~K and $T_2 \approx$ 41.8~K.
This is reminiscent of the double-peak
found in \HFO \cite{bhattacharjee2002heat}.
Measurements on oriented single crystal samples might be
essential to closely track the twin-peaks.
With the application of external magnetic field, the phase transitions
at $T_\mathrm{N1}$ and $T_\mathrm{N2}$ are observed to diminish.
The ZFC magnetization is presented in figure~\ref{fig:mt} (c)
where we can see that both the transitions vanish
with the application of magnetic field of 70~kOe.
An enlarged view of the region, 5 K $\textendash$ 80 K, around the 
transition is provided in the inset of the panel where the
magnetic field is varied from 100~Oe to 70~kOe.
These results support the predominant antiferromagnetic nature of the
observed phase transitions in \HFCO.
\\
In figure~\ref{fig:acchi} (a), the real part of AC susceptibility $\chi'(T)$
of \HFCO\ measured with applied frequencies 0.3~Hz $\textendash$ 999~Hz
in the temperature range 5~K $\textendash$ 100~K is presented.
While the phase transition at $T_\mathrm{N2}$ is captured in $\chi'(T)$,
no frequency dispersion is observed.
The double peak that was observed in the $d\chi/dT$ plots is not present
in $\chi'(T)$. AC susceptibility of \HFCO\ recorded with an applied frequency of
333~Hz and amplitude of 2.5~Oe, superposed with different DC fields
(100~Oe $\textendash$ 70~kOe) is presented in the panel (b) of figure~2.
In presence of a magnetic field, the peak at $T_\mathrm{N2}$ is found to be 
diminished. With the application of higher fields, 40~kOe and 70~kOe, it shifts 
to low temperatures and displays a broad peak. 
It could happen that the application of magnetic field suppressed one of the peaks
at double transitions and enhanced the other.
Or, in other words, one of the magnetic phases is being
converted in to the other through a continuous rotation of
spins \cite{belov1976spin,gorodetsky1969spin}.
\\
In figure~\ref{fig:acchi} (c), the isothermal magnetization of
\HFCO\ is presented for temperatures in the range 5~K $\textendash$ 100~K.
No sign of ferromagnetic saturation is seen attained at 5~K with the application of 70~kOe.
Also, there is no indication of opening of a ferromagnetic loop.
In Table~\ref{tab:spins}, the possible combinations of the spins of Ho, Co and Fe in
low spin (LS), intermediate spin (IS) and high spin (HS) states are given.
The experimental saturation moment of \HFCO\ at 5~K and 70~kOe was estimated
from the law of approach to saturation as 11.2(4)~$\mu_\mathrm{\tiny B}/$fu.
Comparing this experimental value with the effective moments resulting from the
different combination in Table~\ref{tab:spins}, it appears that \HFCO\ has LS Co$^{3+}$,
HS Fe$^{3+}$ and paramagnetic Ho$^{3+}$ spins. 
\subsection{\label{spheat}Specific heat}
\begin{figure}[!t]
	\centering
	\includegraphics[scale=0.43]{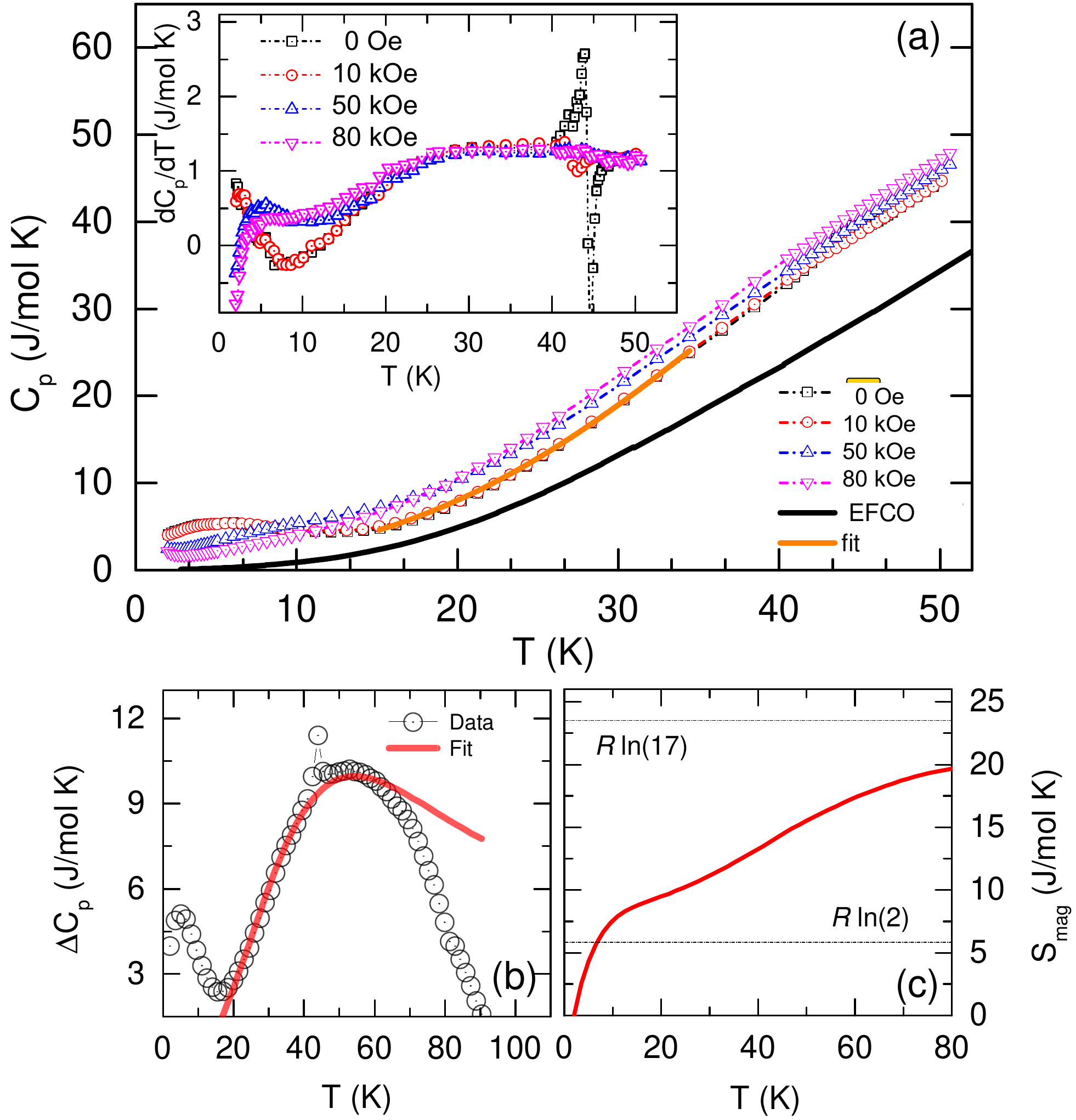}
	\caption{(color online) (a) The specific heat of \HFCO\ in zero and applied fields of $H$ = 10~kOe, 50~kOe and 80~kOe in the temperature range 2~K - 50~K. The phase transition at $T_N$ is clearly seen at $\approx$ 45~K and the rare earth magnetism as a broad peak at $\approx$ 5~K. With the application of the field, the peak at 45~K is suppressed. The low temperature fit (15~K \textendash 35~K) is shown as an orange solid line. The inset is a plot of $dC_p/dT$ versus $T$. (b) The $\Delta C_p$ is plotted along with a fit using a Schottky model of 7 crystal field levels. (c) The magnetic entropy $S_\mathrm{mag}$ estimated from $\Delta C_p$. \label{fig:spheat1}}
\end{figure}

The total specific heat $C_p(T)$ of \HFCO\ in 0~Oe, 10~kOe, 30~kOe, 50~kOe and 80~kOe
are presented in the main panel of figure~\ref{fig:spheat1} (a).
A broad peak centred at approximately 5~K arises from the
magnetism of rare earth where the absence of a sharp $\lambda$-type anomaly
suggests that the magnetic order in the Ho-sublattice might be of short-range
type. We denote this anomaly as $T_\mathrm{Ho}$.
A peak is observed at $T_\mathrm{N1}$ (not shown) in the specific heat
where the paramagnetic to antiferromagnetic phase
transition happens.
At $T_\mathrm{N2} \approx$ 45~K, the $C_p(T)$ displays a peak which
arises from the spin reorientation of the Fe moments.
The inset of figure~\ref{fig:spheat1}(a) provides a plot of $dC_p/dT$ versus
temperature, highlighting the magnetic phase transitions at
$T_\mathrm{N1}$ and $T_\mathrm{N2}$.
The zero field specific heat in the inset confirms that the anomaly at
$T_\mathrm{N2}$ indeed has a double-peak structure as was
observed in the DC magnetic susceptibility.
Also evident from the inset is the fact that the peak in $C_p(T)$
at 45~K is diminished with the application of external magnetic field
$\textendash$ an indication that antiferromagnetic
interactions are important in this compound.
On the other hand, the specific heat at $T_\mathrm{Ho}$ is enhanced
with the application of magnetic field, especially, 50~kOe and 80~kOe.
The low temperature region (from 15~K $\textendash$ 30~K) of the specific heat of \HFCO\
 is fitted using the expression
$C_p(T)$ = $\beta T^3$ + $\alpha T^5$ + $\gamma T$, where the terms 
containing $\alpha$ and $\beta$ correspond to phonon contribution while 
the term with $\gamma$ describes the electronic part. The specific heat 
data in zero and applied fields are fitted to this
expression and the resulting fit parameters are shown in Table~\ref{tab:cp}.
The fit so obtained for the 10~kOe specific heat data
is shown in figure~\ref{fig:spheat1} (c) as a solid line.
Using the value of $\beta$, the
Debye temperature is then estimated as
$\theta_\mathrm{D}$ = $\left(\frac{12p\pi^4R}{5\beta}\right)^{1/3}$
where $p$ is the number of atoms in unit cell and $R$ 
is the universal gas constant. In the low temperature region,
$\theta_\mathrm{D}$ = 138~K is obtained for \HFCO  .
\\
\indent
The specific heat of \EFCO\ is plotted in the main panel of
figure~\ref{fig:spheat1} (a) as a black solid line.
Using this as the non-magnetic analogue (Eu$^{3+}$ being diamagnetic) and subtracting it
from the specific heat of \HFCO\, we recover the magnetic
specific heat corresponding to Ho$^{3+}$.
This quantity, $\Delta C_p$ = ($C_p$(Ho$_2$FeCoO$_6$) $\textendash$ $C_p$(Eu$_2$FeCoO$_6$))
for the temperature range 2 K to 100 K is plotted in figure~\ref{fig:spheat1} (b).
In addition to the broad peak at $T_\mathrm{Ho}$
due to the short-range magnetic order in the Ho sub-lattice,
$\Delta C_p$ also recovers a Schottky-like peak for \HFCO\
centred at around 50~K ($T_\mathrm{Sch}$) which is
similar to the case of \HFO \cite{bhattacharjee2002heat}.
The spin reorientation of the Fe spins appears as a sharp peak at 45~K.
We analyze the Schottky-like peak of $\Delta C_p$ by using a
multi-level crystal field model similar to that used in the case of
\HFO \cite{bhattacharjee2002heat}.
The fit obtained in the case of \HFCO\ is shown
as a solid line in figure~\ref{fig:spheat1} (b).
An 8 level crystal field model was found to roughly account for the Schottky-like peak
centred at $T_\mathrm{Sch}$.
The temperature ranges 15~K $\textendash$ 40~K and
50~K $\textendash$ 60~K were used for the fit, thereby avoiding the region near the peak
at $T_\mathrm{N2}$. From the fit we estimate crystal field energy levels
as 0~K, 79~K, 160~K (doublet), 165~K, 672~K (doublet) and 683 K.
Though a detailed estimation of the crystal field levels would require
neutron inelastic scattering experiments,
the values that we provide here are from a first-principles
curve fit assuming the values for that of \HFO\ as a starting point.
In the case of \HFO\, the low symmetry of the Ho position
leads to the assumption of 17 independent crystal field levels.
Bhattacharjee {\em et al.,} \cite{banerjee1964generalised} 
reports 0~K (doublet),  110~K (degeneracy = 2),
230~K (4), 330~K (2) and 700~K (7) through a trial-and-error
curve fit.
\\
The magnetic entropy of \HFCO\ is estimated from $\Delta C_p$
as $S_\mathrm{mag}$ = $\int (\Delta C_p/T)dT$ and is plotted
in figure~\ref{fig:spheat1} (c).
Two horizontal black lines in the graph indicate the location
of $R$~ln(2) and $R$~ln(17).
It can be seen that the transition at \THo\ recovers
entropy close to $R$~ln(2) suggesting a doublet, which
was the case in \HFO \cite{bhattacharjee2002heat}.
Thus, combining ac and dc magnetization and specific heat,
we identify two magnetic phase transitions in \HFCO\
at 250~K and at 45~K and also signature of short-range magnetic
order of the rare earth at 5~K.
The transition at 45~K is identified as a spin reorientation
transition. In order to confirm this and to obtain the details of the
crystal and magnetic structure, we proceed now to analyze the
neutron diffraction data obtained on \HFCO.
\subsection{Neutron diffraction}
\begin{figure}[!t]
	\centering
	\includegraphics[scale=0.25]{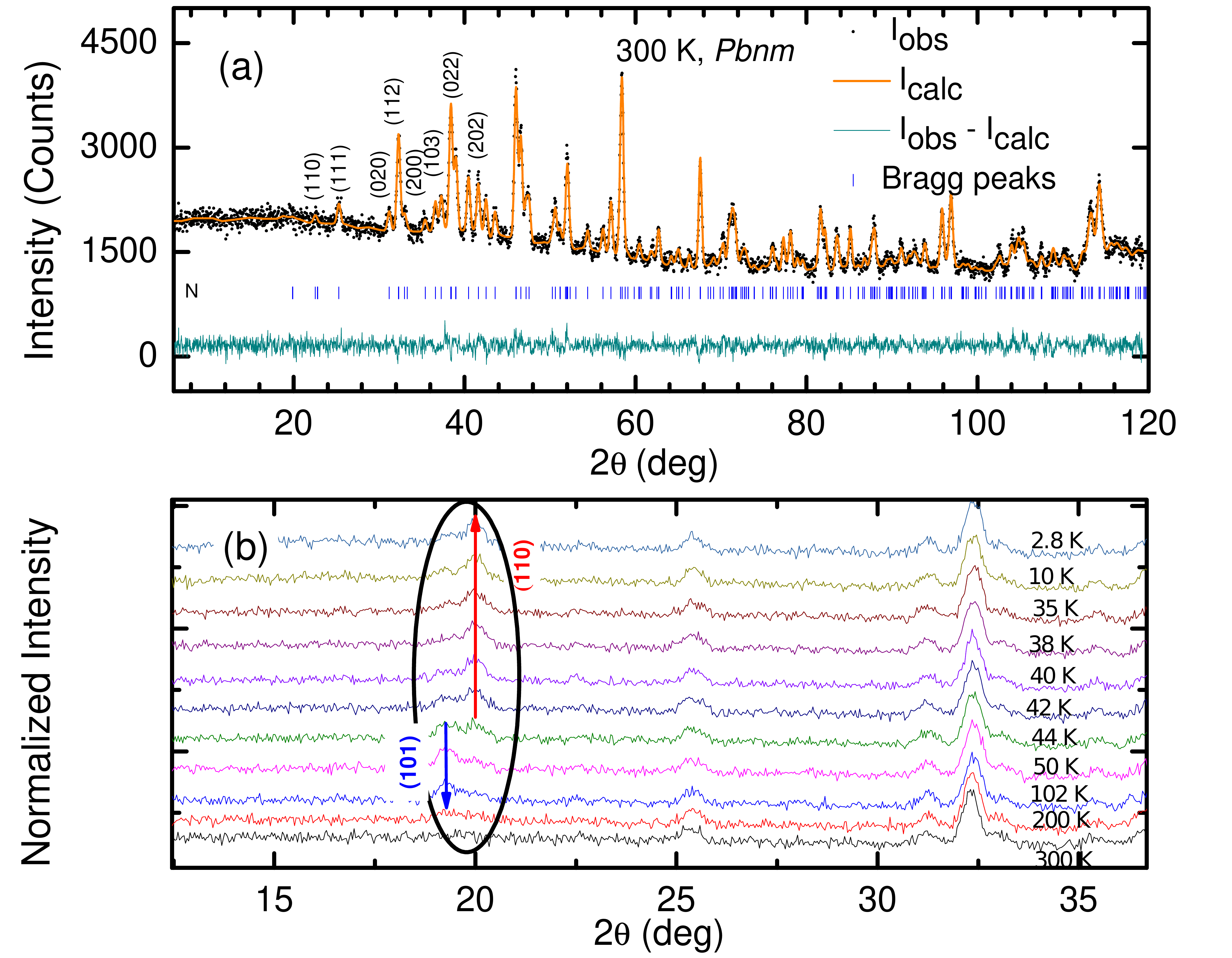}
	\\
	\includegraphics[scale=0.25]{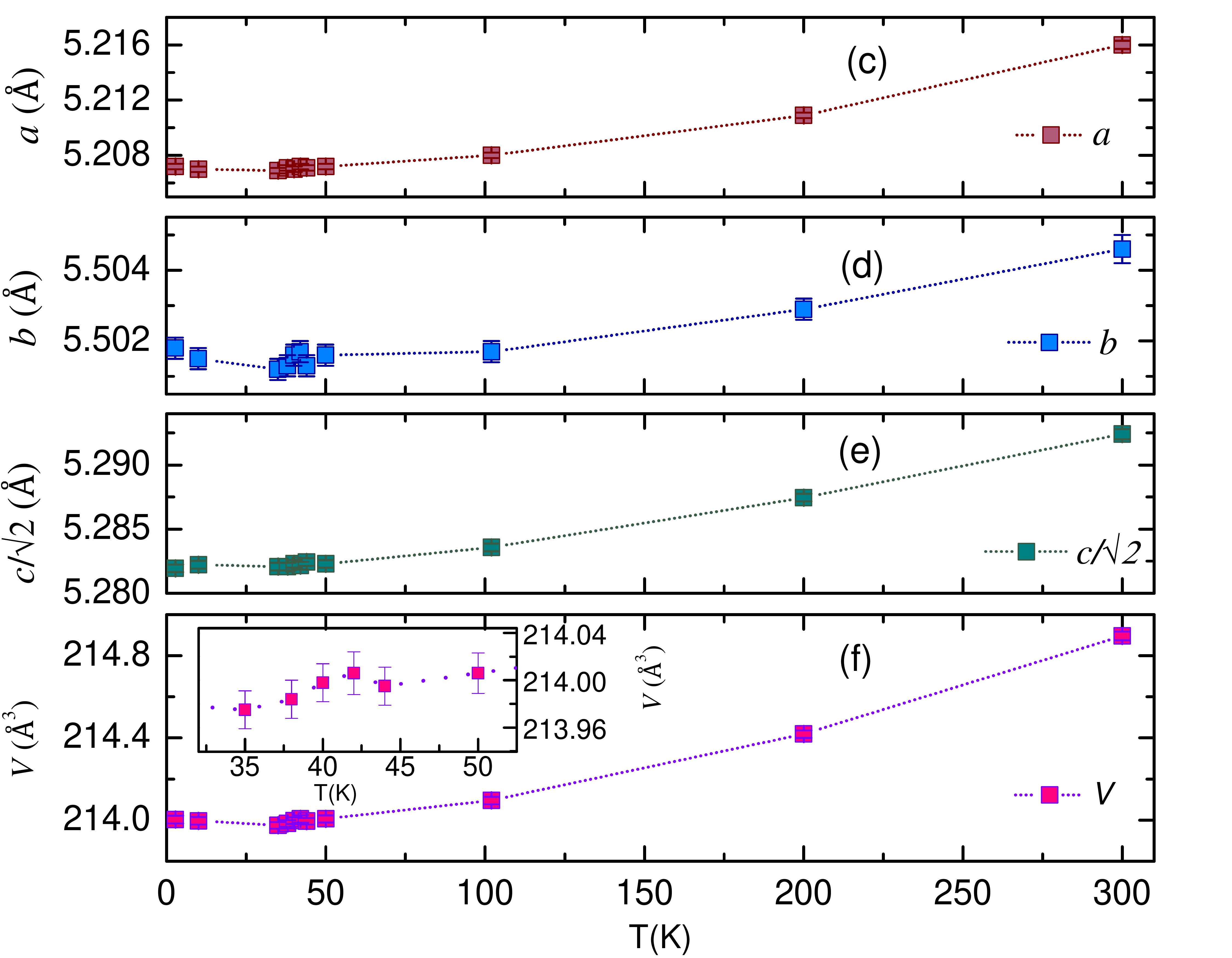}
	\caption{(color online) (a) The neutron diffraction data of \HFCO\ at 300~K.
		The experimental data are shown in dark circles and the Rietveld fits assuming $Pbnm$ space group are shown as orange lines.
		(b) The temperature evolution of the neutron diffraction patterns from 300~K to 2.8~K.
		The intensity of the (101) peak increases from 44~K to 200~K; below 44~K the intensity shifts to the (110) peak. (c)-(e) The temperature dependence of lattice parameters a, b and c respectively. (f) The main panel shows temperature evolution of unit cell volume V, inset shows the anomaly observed around T$_{N2}$. }
	\label{fig:npd1}
\end{figure}

\begin{figure}[!t]
	\centering
	\includegraphics[scale=0.28]{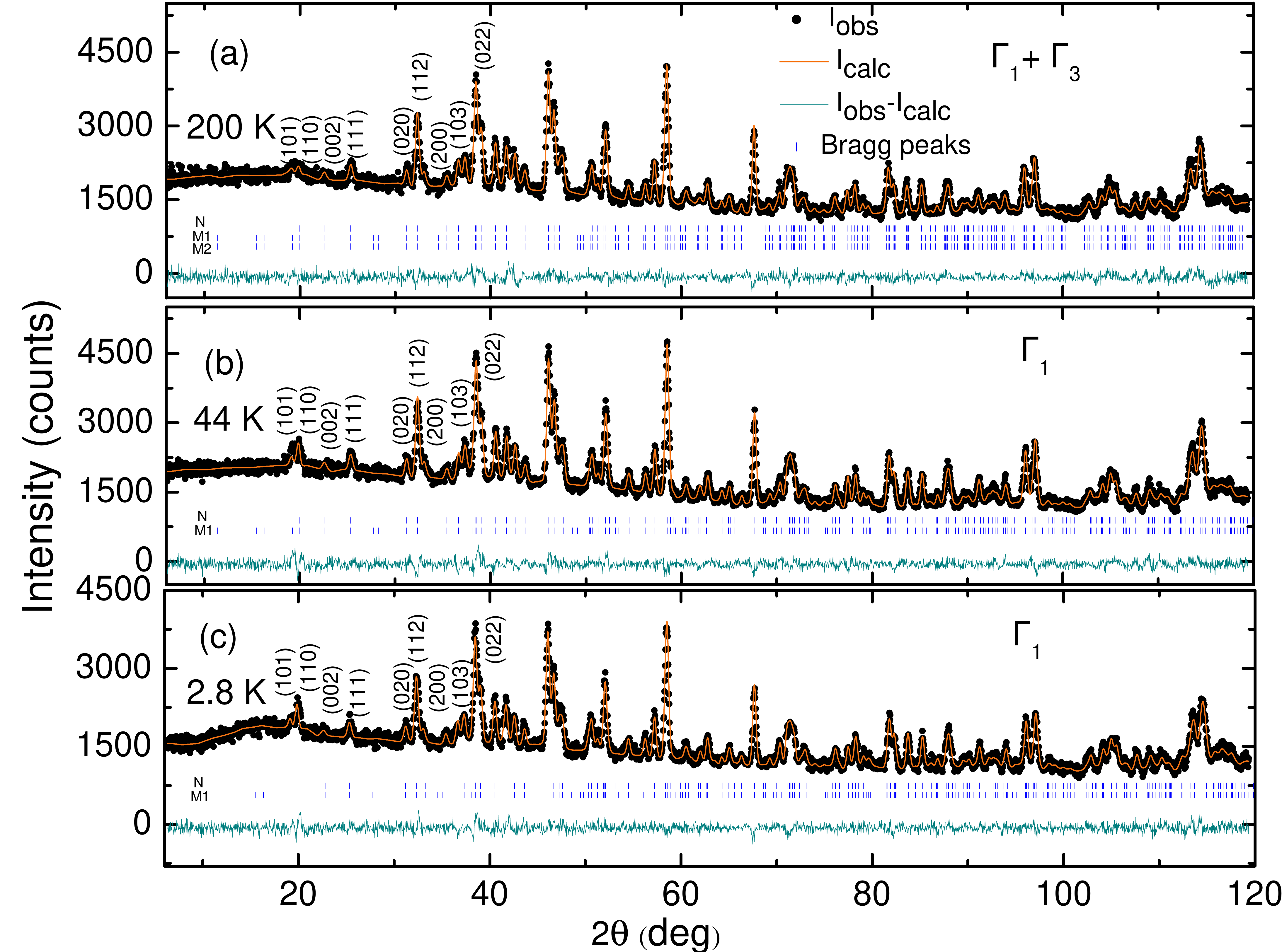}
	\includegraphics[scale=0.28]{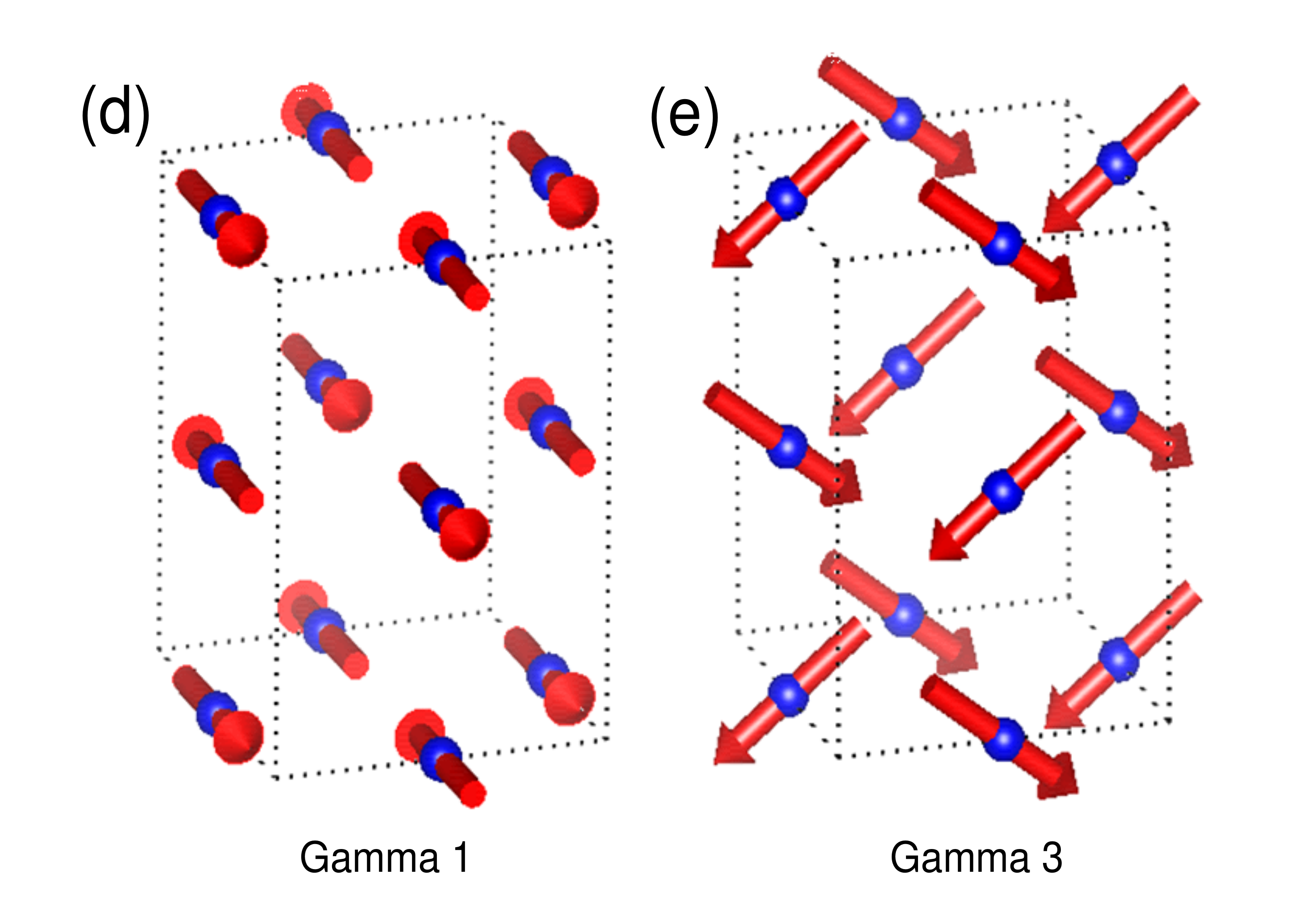}
	\includegraphics[scale=0.28]{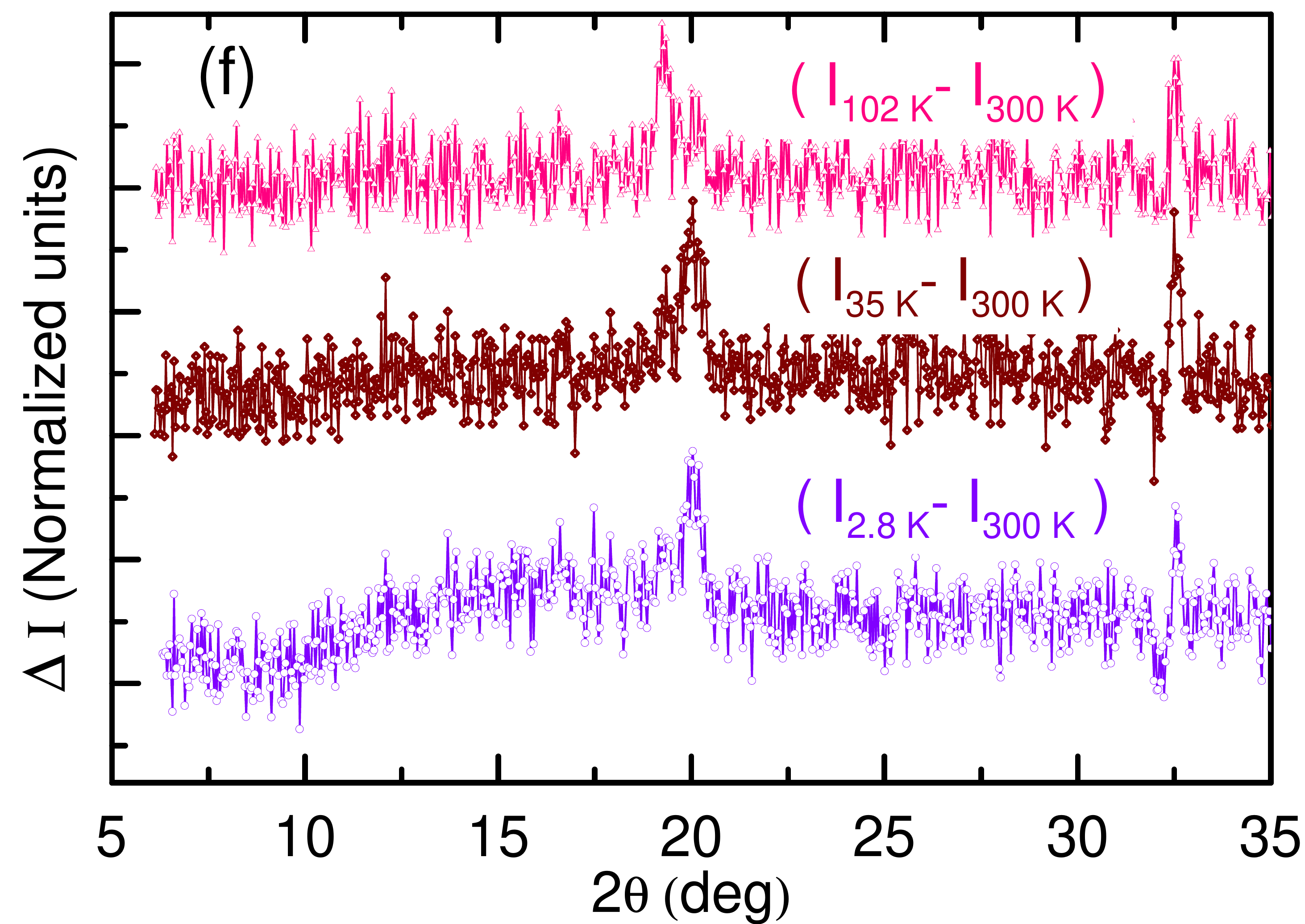}
		\caption{(colour online) (a) The neutron diffraction intensity of \HFCO\ at (a) 200~K (b) 44~K and (c) 2.8~K presented along with the results of Rietveld refinement. At 200~K the magnetic structure consists of the mixed phase ($\Gamma_1$ + $\Gamma_3$). At 44~K, the mixed phase transforms to a single phase $\Gamma_1$ and this phase remains down till 2.8~K. The magnetic structures of (d) $\Gamma_1$ and (e) $\Gamma_3$, visualized using VESTA \cite{momma2011vesta}. (f) The diffuse scattering signal obtained at 102~K, 35~K and 2.8~K by subtracting the 300~K data.}
	\label{fig:npd2}
\end{figure}

The neutron diffraction pattern of \HFCO\ obtained at 300~K
is shown in figure~\ref{fig:npd1} (a).
Our first attempt was to understand the crystal structure,
especially the degree of ordering of Fe and Co.
Generally, the double perovskites crystallize in ordered arrangement for Fe and Co and subsequently the crystal structure is described in the monoclinic space group $P2_1/n$ where Fe and Co occupy the $2c$ and $2d$ Wyckoff positions, respectively.
If the cations are disordered, they occupy a common $4b$ position 
and can be described using the space group $Pbnm$.
The difference in coherent neutron scattering lengths of Fe (9.45~fm)
and Co (2.49~fm) would allow the estimation of the degree of cation ordering,
if studied using neutron scattering.
Hence, our first attempt was to refine the diffraction data in
the cation-ordered $P2_1/n$ space group.
However, if the cations are ordered in the double perovskite structure,
it leads to the appearance of (011) peak in the diffracted intensity \cite{baron2011effect}.
The absence of such a peak in the neutron diffraction data of \HFCO\ suggested
that the degree of ordering is very low in this compound.
Hence, we used $Pbnm$ as the structural model for \HFCO\ where
Fe and Co are randomly occupied on $6c$ crystallographic position.
The Rietveld fit to the 300~K-data is presented in
figure~\ref{fig:npd1} (a) as a solid line.
The refined atomic positions  at 300~K obtained via
the refinement are given in Table~\ref{tab:str}. The temperature dependence of lattice parameters and the unit cell volume~(refer Table \ref{tab:abc}) is presented in panels (c),(d),(e) and (f) of figure~\ref{fig:npd1}. Thermal expansion of the unit cell volume is observed with temperature. The anomaly present in the magnetic data referring to the spin reorientation transition is also reflected in the thermal evolution of unit cell volume. This indicates a close connection between the spin reorientation and the lattice effects in this material.\\
From the magnetisation data presented in Section~\ref{mag},
we know that a magnetic phase
transition from paramagnetic to an antiferromagnetic phase
occurs at $T_\mathrm{N1}\approx$ 250~K.
Hence, below $T_\mathrm{N1}$, the neutron diffraction data is
refined by adding a magnetic phase also to the nuclear structure.
The emergence of the magnetic long-range order is evident
in the neutron diffraction data from the development of
diffracted intensity of the (110) peak that builds up from
200~K down towards 44~K, see figure~\ref{fig:npd1} (b).
Below 44~K, the intensity of the (101) peak diminishes while
that of (110) is enhanced.
This gives the first indication that \HFCO\ undergoes a change
in magnetic structure as a function of temperature.
In order to analyse the magnetic structure of \HFCO\ from the diffraction
pattern at 200~K, we determined the propagation vector
from profile fits to the magnetic peak
at (011) and subsequently using the $k$-search utility in FullProf.
The irreducible representations (IR) of symmetry allowed magnetic structures
for $Pbnm$ space group were obtained using the softwares, BasIreps of FullProf suite \cite{rodriguez2011basireps},\cite{ritter2011neutrons}  and SARA$h$ \cite{sarah}.
Four IR's were suggested - $\Gamma_1$, $\Gamma_3$, $\Gamma_5$
and $\Gamma_7$.
Each of the four magnetic representations was tried in a
trial-and-error method to do a Rietveld fit for the 200~K data.
We found that the representations $\Gamma_1$ and $\Gamma_3$
described the data well. However the best fit to the data
was obtained for the mixed magnetic phase ($\Gamma_1 + \Gamma_3$).
This result also supports the preliminary deduction from the
thermal hysteresis seen in the dc magnetization that the compound
\HFCO\ consists of mixed magnetic structures in the temperature range
from 200~K to at least 45~K ($T_\mathrm{N2}$).
In fact, the refinement of the diffraction data at low
temperatures show that the mixed magnetic phase exists in the
entire temperature region of 200~K to $T_\mathrm{N2} \approx$ 45~K
where the spin reorientation transition occurs.
Below $T_\mathrm{N2}$, only one magnetic phase exists - $\Gamma_1$.
Thus the spin reorientation transition in \HFCO\ is characterized
by the transformation from ($\Gamma_1$ + $\Gamma_3$) to $\Gamma_1$.
It is seen in the inset of figure~\ref{fig:mt} (a)
that the thermal hysteresis also terminates at $T_\mathrm{N2}$.
 At 2.8~K, we analysed the diffraction data to include
the magnetic long-range order in the Ho sub-lattice also. However,
adding a magnetic phase to Ho in the refinement did not improve the
fit or did not better the $\chi^2$ of fit. Moreover, from the
magnetization and the specific heat data no sign of long range
order was suggested for Ho.
The Rietveld refined neutron diffraction data at
200~K, 45~K and 2.8~K are shown in figure~\ref{fig:npd2} (a), (b)
and (c) respectively and the magnetic structure formed by the Fe/Co moments
are shown in (d) and (e). The magnetic moment values obtained at temperatures below 44~K are presented in right most column of table~\ref{tab:abc}. The existence of mixed magnetic phases in the sample from 44~K made it difficult to estimate the accurate moment values for the experimented temperatures at and above 44~K. In addition, the neutron diffraction data
below 10~K show significant enhancement of the intensity in the
low-angle region as a broad peak. The diffracted intensity 
is plotted in figure~\ref{fig:npd2} (f) to show the difference patterns
(102~K $\textendash$ 300~K), (35~K $\textendash$ 300~K) and (2.8~K $\textendash$ 300~K). 
It can be seen that
at 2.8~K, broad diffuse scattering emerges in the
difference pattern. This is a clear indication of the
short-range diffuse magnetic scattering from Ho$^{3+}$.
Hence, we affirm that the Ho$^{3+}$ moment does not order
completely long range in \HFCO\ down to 2.8~K.
\section{Conclusions}
The double perovskite compound \HFCO\ displays two magnetic phase
transitions, one at $T_\mathrm{N1} \approx$ 250~K and the second
at $T_\mathrm{N2} \approx$ 45~K. The first transition marks the
paramagnetic to antiferromagnetic mixed-phase region where the
magnetic structure is described by the representation ($\Gamma_1$ + $\Gamma_3$)
which is antiferromagnetic. After coexisting for a large temperature range, 
the mixed magnetic structure undergoes a spin reorientation 
and transforms to pure $\Gamma_1$ phase at $T_\mathrm{N2}$.
Though the signature of long-range magnetic order is clear in the
transition metal lattice, only 
short-range magnetic order exists in the Ho$^{3+}$-lattice.
\section*{Acknowledgements}
HGR acknowledges Prof. A. M. Strydom and Dr. K. Ramesh Kumar (Highly Correlated Matter Research Group, University of Johannesburg, South Africa) for zero field HC measurement on \EFCO, Dr. Radhika V. Nair (IIT Madras) for constant help and support provided during sample synthesis and Mr. P. Seenivasan (IIT Madras) for the help during SVSM experiments. 
\newpage
\section*{References}
\bibliography{RefHFCOJPCM}
\bibliographystyle{iopart-num}
\newpage
\begin{center}
	\begin{table}[!b]
		\setlength{\tabcolsep}{10pt}
		\caption{The effective magnetic moment values of \HFCO\ (calculated theoretically considering the paramagnetic moments of the constituent cations viz., spin only moments of Fe, and Co and spin-orbit coupling of Ho), assuming different valence states (V) and  spin states (SS) for Ho, Co and Fe. The magnetic moment values are given in the units of $\mu_\mathrm{B}$. The total moment $\mu_\mathrm{tot}$ is estimated as $\sqrt{\mu^2_\mathrm{Ho} + 0.5 (\mu^2_\mathrm{Fe} + \mu^2_\mathrm{Co})}$.  (LS: low spin, IS: intermediate spin and HS: high spin).  \label{tab:spins}}
		\centering
		\begin{tabular}[h]{cccc}
			
			\hline
			Ho$^{3+}$        &   Co                                        & Fe     &  $\mu_\mathrm{tot.}$    \\
			($\mu_\mathrm{Ho}$)  & (V, SS, $\mu_\mathrm{Co}$)     & (V, SS, $\mu_\mathrm{Fe}$)   & ($\mu_\mathrm{B}$) \\ \hline\hline
			10.6                 & +3, LS, 0                                   & +3, LS, 1.73             & 10.67  \\
			&                                             & +3, HS, 5.92             & 11.39  \\
			&                                             & +4, LS, 2.82             & 10.79  \\
			&                                             & +4, HS, 4.89             & 11.15  \\  \hline
			10.6                 & +3, IS, 2.82                                & +3, LS, 1.73             & 10.86  \\
			&                                             & +3, HS, 5.92             & 11.56  \\
			&                                             & +4, LS, 2.82             & 12.01  \\
			&                                             & +4, HS, 4.89             & 11.32  \\  \hline
			10.6                 & +3, HS, 4.89                                & +3, LS, 1.73             & 11.22  \\
			&                                             & +3, HS, 5.92             & 11.91  \\
			&                                             & +4, LS, 2.82             & 11.33  \\
			&                                             & +4, HS, 4.89             & 11.67  \\  \hline
			10.6                 & +4, LS, 1.73                                & +3, LS, 1.73             & 10.74  \\
			&                                             & +3, HS, 5.92             & 11.46  \\
			&                                             & +4, LS, 2.82             & 11.33  \\
			&                                             & +4, HS, 4.89             & 11.22  \\  \hline
			10.6                 & +4, HS, 5.92                                & +3, LS, 1.73             & 11.46  \\
			&                                             & +3, HS, 5.92             & 12.14  \\
			&                                             & +4, LS, 2.82             & 11.57  \\
			&                                             & +4, HS, 4.89             & 11.91  \\  \hline\hline

		\end{tabular}
	\end{table}
\end{center}
\clearpage\newpage
\begin{center}
	\begin{table}[!b]
		\setlength{\tabcolsep}{4pt}
		\caption{The fit parameters  obtained by analysing the low temperature specific heat of \HFCO in the temperature range 15~K $\textendash$ 30~K.
			The Sommerfeld coefficient $\gamma$ and the Debye temperature $\theta_\mathrm{D}$ estimated from the fit are shown.  
			\label{tab:cp}}
		\centering
		\begin{tabular}{ccccc}  
			\hline
			H         &   $\alpha$                   & $\beta$                  & $\gamma$     &   $\theta_\mathrm{D}$ \\
			(T)       &   (10$^{-8}$ J/mol K$^{-6}$)   & (10$^{-4}$J/mol K$^{-4}$)  & (J/mol K$^{-2}$)    &           (K)             \\ \hline\hline
			0         & -19.0                        & 72.9                     & 0.128        &          138 \\
			1         & -14.8                        & 6.57                     & 0.160        &           -       \\
			5         & -8.89                        & 4.74                     & 0.343        &           -       \\
			8         & -23.6                        & 7.36                     & 0.271        &           -       \\ \hline\hline
		\end{tabular}
		
	\end{table}
\end{center}
\clearpage\newpage
\begin{center}
	\begin{table}[!b]
		\setlength{\tabcolsep}{10pt}
		\caption{The fractional coordinates of atoms in \HFCO\ at 300~K according to $Pbnm$ space group setting.
			\label{tab:str}
			}
		\centering
		\begin{tabular}{l|llll}   \hline
			& $x$          & $y$             & $z$                  & $\mathrm{B_{iso}}$               \\ 
			\hline\hline
			Ho        & 0.9862(9)       & 0.0689(5)          & 0.25               &   0.0073             \\
			Co/Fe        & 0.5            & 0             & 0                    &   0.0085             \\
			Fe/Co        & 0.5          & 0               & 0                    &   0.0085             \\
			O(1)      & 0.0995(9)       & 0.4724(9)          & 0.25               &    0.0136            \\
			O(2)      & 0.6943(8)       & 0.2994(8)          & 0.0526(5)               & 0.0147               \\ \hline\hline
		\end{tabular}
	\end{table}
\end{center}
\clearpage\newpage
\begin{center}
	\begin{table}[!b]
		\setlength{\tabcolsep}{10pt}
		\caption{The lattice parameters and the magnetic moments extracted using Rietveld refinement of neutron diffraction data. 
			\label{tab:abc}
			}
		\centering
		\begin{tabular}{llllll}   \hline
		 \textit{T}	         &      \textit{a}             &    \textit{b}       &    \textit{c}           &    \textit{V}        &  Magnetic moment($\Gamma_1$)       \\ 
		 (K)          &   (\AA{})          &  (\AA{})   &  (\AA{})       &  (\AA{}$^3$) &  $\mu_\mathrm{B}$                  \\
			\hline\hline
		2.8 & 5.2071(2) & 5.5020(3) & 7.4696(4) & 214.000(19) & 3.2(1) \\
		10  & 5.2070(2) & 5.5015(3) & 7.4703(3) & 213.997(18) &3.5(1)\\
		35  & 5.2069(2) & 5.5013(3) & 7.4702(3) & 213.981(16) &3.0(1)\\
		38  & 5.2071(2) & 5.5014(3) & 7.4700(3) & 213.988(16) &3.4(1)\\
		40  & 5.2070(2) & 5.5017(3) & 7.4703(3) & 214.004(18) &3.2(1)\\
		42  & 5.2072(2) & 5.5017(3) & 7.4701(3) & 214.007(18) &3.7(1)\\
		44  & 5.2072(2) & 5.5012(3) & 7.4704(3) & 213.995(16) &2.7(2)\\
		50  & 5.2072(2) & 5.5016(3) & 7.4704(3) & 214.011(17)                     
		                                        &\\
		102 & 5.2080(2) & 5.5017(3) & 7.4721(3) & 214.097(18) &\\
		200 & 5.2110(2) & 5.5028(3) & 7.4775(3) & 214.419(18) &\\
		300 & 5.2159(1) & 5.5048(9) & 7.4846(6) & 214.897(21) &\\    
 
		 \hline\hline
		\end{tabular}
	\end{table}
\end{center}
\clearpage\newpage
\clearpage\newpage
\clearpage\newpage
\clearpage\newpage

\clearpage\newpage

\end{document}